# High Power Polarized Positron Source

Alexander Mikhailichenko
*Cornell U., Ithaca NY 14853*

*Abstract.* We discuss the basics of polarized positron production by low energy polarized electrons. Efficiency of conversion ~0.1-1% might be interesting for CEBAF facility and ILC as well.

## 1. OVERVIEW

Positrons could be obtained by two ways: first way–by Beta-decay, second way–by (polarized) gammas in field of nuclei.

Creation of a pair by photons is described in [1], [2]. In turn there are three ways to create polarized gammas, suitable for further positron production in practical amounts: (a) radiation in helical field arranged with help of EM wave, radiation in static field of helical undulator and by radiation due to channeling in helical crystal [3], (b) by scattering on laser radiation as some particular case of EM wave [4]. (c) by Bremstrahlung of initial electron having longitudinal polarization [5]. The process (c) could be recommended for CEBAF.

For ILC polarized positrons are created by gammas generated by the main beam in helical undulator $K<1$, $L\sim150m$, $\lambda_u = 1cm$, (a). We also mentioned the possibility to use (c) for ILC [14].

## 2. EFFICIENCY OF POSITRON PRODUCTION

Polarized positron production is a two-stage process: at first stage circularly polarized electrons generate longitudinally polarized gammas, and then, at the second stage, these gammas are converted into longitudinally polarized positrons in the same media. At the high edge of spectra, positrons have longitudinal polarization. Polarization could be enhanced by selection of positrons by theirs energy: higher energy linked with higher polarization. This procedure stays in line with limiting energy acceptance of CEBAF SRF structure.

So the components of Positron Source suitable for CEBAF include (a) source of (Polarized) electrons 10-200MeV (ILC source has ~48 µA ; polarized source operating at CEBAF gives 100 µA at present; plans to have 200 µA ,DULY R Inc [12]); (b) High power Target; (c) Beam collection system. Protection and shielding is last, but not least in this schedule.

Reaction of neutron photo-production has a threshold, which depends on media. For Tungsten, the gammas with energy $E< 6.19$ *MeV* can not generate neutrons [7], [8]; for reaction of ($\gamma$, 2n) the threshold is 13.6 *MeV* (W). For some materials thresholds could be found in [11] (see references there). One possible scenario–to use for conversion electrons having energy below the threshold could not be recognized as a productive one as the efficiency of positron production by gammas at 5 *MeV* compared with 50 *MeV* drop at least 10 times plus efficiency of creation of gammas drop $\sim \gamma^2$ i.e. 100 times.

One can find calculations of efficiency for 50 MeV electrons in [5], so here we omit details due to lack of space. The total number of positron created by each electron finally comes to [5]

$$\Delta N_+ \cong \frac{d^2}{2Ln(183Z^{-1/3})} \begin{cases} \frac{19}{15} \cdot (1-\xi_+)^{5/2}, 0.89 < \xi_+ \leq 1 \\ 0.015 + \frac{31}{40}(0.089 - \xi_+)^2, 0.11 \leq \xi_+ \leq 0.89 \end{cases} \quad (1)$$

where $\xi_+ = E_+ / E_+^{\max}$, thickness of target *d* measured in units of radiation length $X_0$. Polarization could be approximated as $\varsigma_+ = \xi_2 \cdot [1 - 2(1-\xi_+)^2]$ [1]. Emittance of the beam is $\sim \varepsilon_x^+, cm \cdot rad \cong 0.084$ [10]. Example: for $\xi_{+\max} = 0.25$, i.e. collection arranged for the positrons in 25% energy interval around maximal energy, *d*=0.3, then $\Delta N_+ = 1.5 \cdot 10^{-2}$, i.e. efficiency 1.5%. Degree of polarization

$$\xi_{e+} = \varsigma_e \cdot 0.91 \cdot 0.87 \cong 0.79 \varsigma_e \to 0.63, i.e.\ 63\%$$

Finally we have a number for efficiency, ~1% which means, that the electron current must be 100 bigger, than the positron one. Now the efficiency of collection must also be taken into account. For narrow energy interval this could be ~20%, coming to the current ratio *Ielectron/Ipositron*~500 So for positron current desired *Ipositron*=1 *micro*Ampere, the electron current must be *Ielectron*= 0.5 mA Target has a thickness~ 0.3 $X_0$ which corresponds to ~0.12*cm*=1.2*mm* for Tungsten or Lead; Titanium will be extremely non effective here.

Energy deposited in a target is $2MeV/(g/cm^2)$ x0.3x$6.8g/cm^2$ =4.1*MeV*. This yields the power deposition in a target ~0.5*mA*x4*MeV*~2 *kW* So the target must be designed for ~5 *kW*. For positron current 0.1 $\mu$A the power comes to moderate 0.5 *kW*. Effective depth at which positrons created is $l \cong <x\theta_x> / <\theta_x^2>$.

## 3. HIGH POWER TARGETS

High power targets could be subdivided into two classes: rotating targets and liquid metal ones. Any of these installations has in purpose a distribution of deposited energy in as big a volume of material as possible.

**3.1 Liquid metal target.** Let us make some estimations for the liquid metal target. Pb/Bi: Liquid at $T_l$=125.9°C, boiling phase at $T_b$=~1500°C, latent heat =860 *kJ/kg*, $\rho$~10*g/cm*$^3$. $^{83}$Bi-$^{82}$Pb alloy composed with 55.51Mass% of Bi and 44.49 Mass% of Pb. Other alloy, Pb-Sn, composed by 61.9 % of Sn has liquid phase at ~183 °C – advantage of Bi-Pb is obvious.

Mercury: Z=80, liquid at $T_l$=-38.87°C, boiling at $T_b$=~357°C. Latent heat =294 *kJ/kg*, $\rho$~13.54 *g/cm*$^3$. One negative property of Mercury, what may strictly influence the choice – is its toxicity. Hg is considered one of most toxic materials; however it could be handled properly. In some installations the Mercury is in use in the turbine circle, instead of water which is what gives assurance of success of its implementation for the purposes of positron production. Usage of Hg as a proton target described in [17].

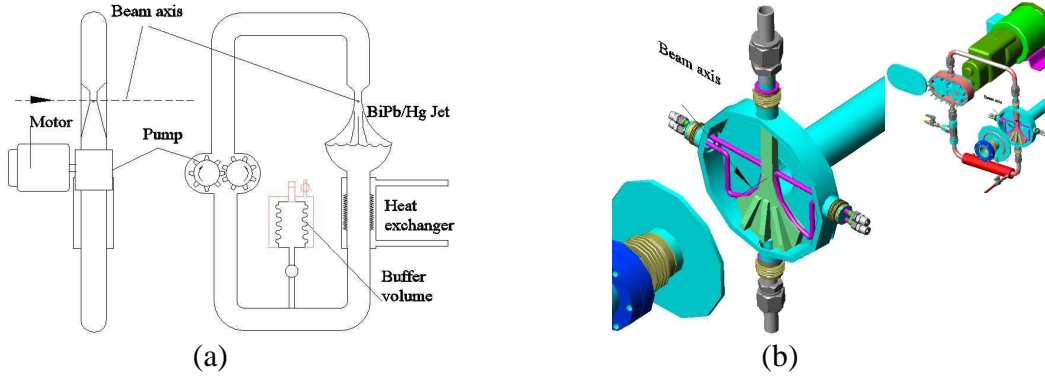

**FIGURE 1.** To avoid damage we suggested that liquid metal jet hits the free liquid metal surface (a) ; isometric view on the target, (b).

Let the jet transverse size be 2*mm*, size along the beam -1.2*mm*, velocity of jet=10*m/sec* so in 1 second the volume of material flowing through the nozzle comes to $V \approx 2.4 \times 10^4$ $mm^3 = 2.4 \times 10^{-2}$ $dm^3$. For average power deposition 5*kW* ($Q$=5 *kJ* per second) the temperature rise of this amount of material comes to

$$T \cong \frac{Q}{mC} = \frac{Q}{\rho VC} = \frac{5}{10 \times 0.024 \times 0.13} \cong 160°C$$

It begins vaporizing, so the latent heat of vaporization needs to taken into account, and it will take $\Delta Q \approx 860 \times 0.024 \times 10 = 206$ *kJ* for Bi-Pb and $\Delta Q \approx 95kJ$ for Hg. So Bi-Pb will remain at ~300 $^o$C and Hg at 160 $^o$C.

The jet chamber could be made from Ti (melt @1668$^o$C) or Niobium (melt @ 2464$^o$C) .As the energy deposited in the windows could be made small, their cooling could be carried by the metal jet itself.
Material for windows are $^4$Be; $^{22}$Ti; Boron Nitride- BN ($^5B^7N$, sublimates @2700$^o$C) and Carbon.
Window could be omitted, but this will require differential pumping and cooled traps.

### 3.2 Rotating target.

Different types of rotating targets have been discussed in literature for a long time [16]. Due to cyclic nature of motion the volume involved in the process is limited. This type of target could be recommended for power deposition below 10 kW. Examples of beam dump for electrons and photons one can find in [11].

## 4. FOCUSING

Short, Iron free solenoidal lens is mostly adequate here. Focal distance *f* of solenoidal lens with longitudinal field at axis $H_\parallel (s)$ can be found from the following expression

$$f^{-1} \cong \frac{GL}{(HR)} = \frac{1}{4(HR)^2} \int H_\parallel^2(s) ds \qquad (2)$$

where $(HR) = pc/300$ stands for magnetic rigidity ~33*kG*×*cm* for 10*MeV* particles. For *f*~2 *cm* Integral comes to ~66$kG^2$×*cm*; For contingency ~4*cm* the field value comes to 4 *kG*. For generation of such field the amount of Ampere-turns required goes to be

$$nJ \cong \frac{H_{\parallel \max} l}{0.4\pi} \cong \frac{4 \times 4}{0.4\pi} \cong 12.7 kA \times turns$$

For the number of turns =10, current in one turn goes to $I_1$~1.3*kA*. Conductor cross-section might be ~ 5x10*mm*$^2$; Cooling carried with de-ionized water. Current density ~26A/*mm*$^2$, which is ordinary, as the maximal achievable in practice is $\leq$ 100 *A*/*mm*$^2$.
No flux concentrator is possible for DC; however the current density is different in accordance with the path length difference, so manipulation with thickness of conductor is possible. Few examples for focusing systems represented below.

### 4.1 E-166 experiment [13].

Experiment E-166 was dedicated to test an undulator-based positron production scheme [4]. Positron collection system includes DC solenoid with Iron yoke running with current density up to 35*A*/*mm*$^2$ delivering focal distance ~4 *cm* for 10*MeV* positrons.

### 4.2 Cornell Positron Source [9].

Modification of the positron source for CESR in 2001 allows for a positron rate ~10$^{11}$/*sec* at 50 *Hz*. Conversion of electrons into positrons is going at 200 MeV. The focusing coil feed by pulsed current up to 3.5kA with pulse duty at halve height ~20 *μs*. It is made as partial flux concentrator. Lot of efforts applies for symmetryzation of field pattern around target itself and in focusing coil, taking into account surroundings. Conversion efficiency with this source ~2.5%, DC power consumption ~2.5 *kW*.

## 5. POSSIBLE SCHEME FOR CEBAF

Possible scheme for CEBAF represented in Fig. 2.

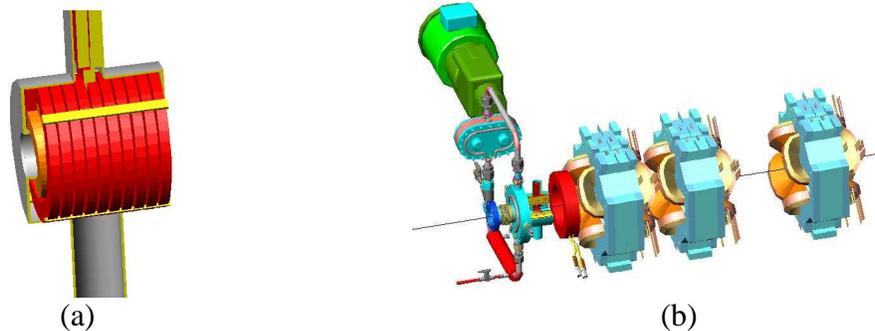

(a)          (b)

**FIGURE 2.** Possible scheme for CEBAF. (a) Two-layer DC solenoid [15]. (b) Liquid metal target, DC solenoid, adjusting coil followed by triplet.

DC compact solenoid, (a) in Fig.2, cooled by de-ionized water. Triplet (b) serves for matching focusing of compact DC solenoid having azimuthtal symmetry and the rest optics (FODO).

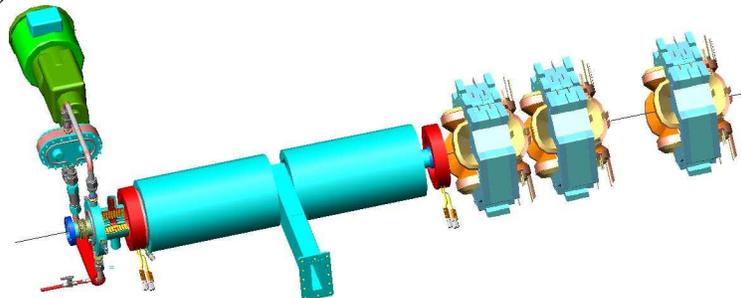

**FIGURE 3.** Addition to the previous scheme is shown here. Al made accelerating structure immersed in solenoid wound with Al conductor.

## 6. SUMMARY

Polarized positrons could be obtained from polarized electrons in a thin target. The source of ~100μA operating at CEBAF could be used for these purposes. One can expect efficiency of electron-positron conversion within 0.1% -1%. Power deposited in a target remains below 5 *kW* for positron current ~1 *μA*; proportionally lower for lower positron current. Target with liquid metal (Bi-Pb) is mostly adequate here. Collection optics as a DC solenoid recommended for CEBAF positron source.

We recommend a design for 20kW of absorption power as a safe margin; the cost will not be raised much.

## 7. REFERENCES

[1] Olsen, L.C. Maximon, "*Photon and Electron Polarization in High-Energy Bremsstrahlung and Pair Production with Screening",* Phys. Rev. 114 (3) (1959 ) 887-904.
[2] H. Bethe, W. Heitler,"*On the Stopping of Fast Particles and on the Creation of Positive Electrons*", Proc. Roy. Soc. A 146 (1934) 83-112.
[3] V.E. Balakin and A.A. Mikhailichenko, "*The Conversion System for Obtaining High Polarized Electrons and Positrons*", Budker Institute of Nuclear Physics, BINP 79-85 (1979).
[4] E.G. Bessonov, "*Some Aspects of the Theory and Technology of the Conversion Systems of Linear Colliders*", Proc. 15$^{th}$ Intl. Conf. on High Energy Accel. (Hamburg, 1992), p. 138.
[5] E.G. Bessonov, A.A. Mikhailichenko, "*A Method of Polarized Positron Beam Production*". June 1996. 3pp., Published in EPAC96, Barcelona, June 9-14, 1996, Proceedings, p.1516-1518.
[6] B. Rossi, "High Energy Particles", N/Y, 1982.
[7] R.Montalbetti, L.Katz, J. Goldemberg, "*Photoneutron Cross Sections*", Phys.Rev. **91**, 659 (1953).
[8] W.P.Swanson, "*Calculation of Neutron Yields Released by Electrons Incident on Selected Materials"*, Health Physics, Vol.35, pp.353-367, 1978.
[9] J. Barley, V. Medjidzade, A. Mikhailichenko, "*New Positron Source for CESR*", CBN-01-19, Oct 2001. 16pp.
[10] V.A. Tajursky, "*Calculation of Electron Conversion into Positrons at 0.2—2GeV"*, Budker INP 76-36, Novosibirsk, 1976.
[11] A.Mikhailichenko, "*Physical Foundations for Design of High Energy Beam Absorbers*", CBN 08-8, Cornell U., LEPP, Ithaca, NY 14853, U.S.A. Available at
http://www.lns.cornell.edu/public/CBN/2008/CBN08-8/CBN08-8.pdf .
[12] D. Yu, M. Lundquist, Y. Luo, A. Smirnov, "*Polarized Electron PWT Photoinjectors*", PESP 2008 at Jefferson Lab., DULY Research Inc. California, USA
[13] G.Alexander *et al*, *"Observation of Polarized Positrons from an Undulator-Based Source",* SLAC-PUB-13145, DESY-08-025, CLNS-08-2023, COCKCROFT-08-03, DCPT-08-24, IPPP-08-12, Mar 6, 2008. 4pp. Published in Phys.Rev.Lett.100:210801, 2008.
[14] A.Mikhailichenko*," ILC Undulator Based Positron Source, Tests and Simulations*", PAC07-WEZAB01, Jun 2007. 5pp. In the Proceedings of Particle Accelerator Conference (PAC 07), Albuquerque, NM, 25-29 Jun 2007, pp 1974-1978.
[15] A.Mikhailichenko*," Collection Optics for ILC Positron Target",* PAC07-WEZAB01, Jun 2007. 5pp. *In the Proceedings of Particle Accelerator Conference (PAC 07), Albuquerque, New Mexico, 25-29 Jun 2007, pp 1974.* Also in *Albuquerque 2007, Particle accelerator* 1974-1978.
http://accelconf.web.cern.ch/AccelConf/p07/PAPERS/WEZAB01.PDF
[16] F. Bulos, H. DeStaebler, S. Ecklund, R. Helm, H. Hoag, H. Le Boutet, H. L. Lynch, R. Miller, K. C. Moffeit, "*Design of a High Yield Position Source*", SLAC -PUB – 3635, April 1985. Published in 1985 in IEEE Trans.Nucl.Sci., vol. 32, pp. 1832-1834.
http://slac.stanford.edu/pubs/slacpubs/3500/slac-pub-3635.pdf .
[17] H.G Kirk *et al* , "*A High-Power Target Experiment",* Particle Accelerator Conference (PAC2005), Knoxville, TN, Proceedings, pp. 3745-3747.